\documentclass{article}
\usepackage{spconf,amsmath,graphicx}
\usepackage{hyperref}
\usepackage{breakurl}
\usepackage{floatrow}
\usepackage{rotating}

\usepackage [english]{babel}
\usepackage [autostyle, english = american]{csquotes}
\MakeOuterQuote{"}

\title{HCU400: An Annotated Dataset for Exploring Aural Phenomenology through Causal Uncertainty}
\name{Ishwarya Ananthabhotla$^{*}$, David B. Ramsay\sthanks{Equal contribution. The authors would like to thank the AI Grant for their financial support of this work.}, and Joseph A. Paradiso}
\address{MIT Media Laboratory, Cambridge, MA.}
\begin{document}
%\ninept
%
\maketitle
\begin{abstract}

\noindent The way we perceive a sound depends on many aspects-- its ecological frequency, acoustic features, typicality, and most notably, its identified source.  In this paper, we present the HCU400: a dataset of 402 sounds ranging from easily identifiable everyday sounds to intentionally obscured artificial ones. It aims to lower the barrier for the study of aural phenomenology as the largest available audio dataset to include an analysis of causal attribution.  Each sample has been annotated with crowd-sourced descriptions, as well as familiarity, imageability, arousal, and valence ratings. We extend existing calculations of causal uncertainty, automating and generalizing them with word embeddings.  Upon analysis we find that individuals will provide less polarized emotion ratings as a sound's source becomes increasingly ambiguous; individual ratings of familiarity and imageability, on the other hand, diverge as uncertainty increases despite a clear negative trend on average.

\end{abstract}
\section{Motivation}
\label{sec:intro}

Despite a substantial body of literature, human auditory processing remains poorly understood.  In 1993, Gaver introduced an ecological model of auditory perception based on the physics of an object in combination with the class of its sound-producing interaction \cite{gaver1993world}.  He suggests that everyday listening focuses on sound sources, while musical listening focuses on acoustic properties of a sound, and that the difference is experiential.  Current research has corroborated this distinction-- studies show that listeners primarily group sounds by category of sound-source, sometimes group sounds by location/context, and only in certain conditions favor groupings by acoustic properties \cite{marcell2000confrontation, gygi2007general}.  Recent work with open-ended sound labeling demonstrates that limited categorization tasks may encourage more detailed descriptions along valence/arousal axes (i.e. for animal sounds) or using acoustic properties (i.e. for mechanical sounds) if sound-source distinctions are too limited for the categorization task \cite{bones2018distinct}.

It has been suggested that non-verbal sounds from a living source are processed differently in the brain than other physical events \cite{lewis2004human}.  Symbolic information tends to underly our characterization of sounds from humans and animals (i.e. yawning, clapping), while acoustic information is relied on for other environmental sounds \cite{giordano2010hearing,aglioti2010representing,pizzamiglio2005separate}.  Furthermore, in \cite{dubois2006cognitive} Dubois et al. demonstrated that, for complex scenes, the perception of pleasant/unpleasantness was attributed to audible evidence of human activity instead of measurable acoustic features. 

It is clear from the above research that any examination of sound phenomenology must start with a thorough characterization of a sound's interpreted cause.  In many cases however, a sound's cause can be ambiguous.  In \cite{ballas1993common} Ballas introduced a measure of causal uncertainty ($H_{cu}$) based on a large set of elicited noun/verb descriptions for 41 everyday sounds: $H_{cui} = \sum_{j}^{n} p_{ij} log_{2} p_{ij}$. (For sound $i$, $p_{ij}$ is the proportion of labels for that sound that fall into category $j$ as decided by experts reviewing the descriptions). He shows a complicated relationship between $H_{cu}$ and the typicality of the sound, its familiarity, the average cognitive delay before an individual is able to produce a label, and the ecological frequency of the sound in his subjects' environment.  $H_{cu}$ was further explored in \cite{lemaitre2010listener} using 96 kitchen sounds.  Lemaitre et al. demonstrated that $H_{cu}$ alters how we classify sounds: with low causal uncertainty, subjects cluster kitchen sounds by their source; otherwise they fall back to acoustic features.  

In this paper, we introduce the HCU400 dataset-- the largest dataset available for studying everyday sound phenomenology.  In this dataset, we include 402 sounds that were chosen to (1) capture common environmental sounds from everyday life, and (2) to fully sample the range of causal uncertainty.  While many of the sounds in our dataset are unambiguous, over 100 of the sounds are modified to intentionally obscure their source-- allowing explicit control of source-dependent effects.  

As part of the dataset, we include high-level emotional features corresponding to each sound's valence and arousal, in line with previous work on affective sound measurement \cite{bradley2007international}.  We also account for features that provide other insights into the mental processing of sound-- familiarity and imageability \cite{schirmer2011perceptual,visualize}.  We explore the basic relationships between all of these features.

Finally, we introduce word embeddings as a clustering technique to extend the original $H_{cu}$, and apply it to the free response labels we gathered for each sound in the dataset.  Deep learning has provided a new tool to represent vast amounts of semantic data in a highly compressed form; these techniques will likely make it possible to model and generalize source-dependent auditory processing phenomena.  The HCU400 represents a first step in that direction.

%\begin{figure}[h]
%	\includegraphics[width=8.5cm]{images/labelingtest2.png}
%    \caption{A screenshot of the interface shown to AMT workers.}
%    \label{interface}
%\end{figure}

\section{Dataset Overview}
\label{sec:format}

The HCU400 dataset consists of 402 sound samples and 3 groups of features: sound sample annotations and associated metadata, audio features, and semantic features. It is available at \url{github.com/mitmedialab/HCU400}. 

\begin{figure*}[h]
\begin{floatrow}
	\includegraphics[width=0.82\textwidth]{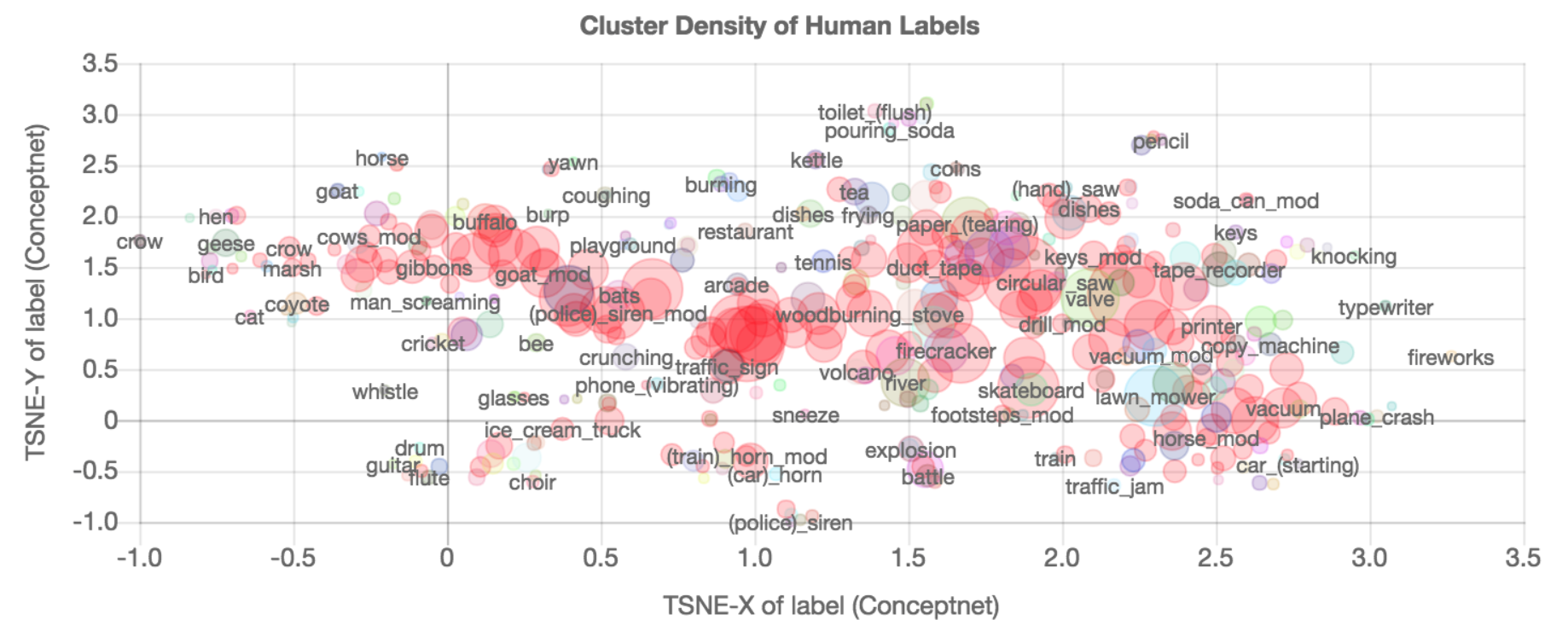}

%   \ffigbox[.4\columnwidth]{
%   \resizebox{\columnwidth}{!}{%

%   \begin{tabular}{|l|l|}
%   \hline
% 	\textbf{Typing} & \textbf{Modified Chair Sliding} \\
%     \hline
%     Cluster Radius = 6.3  & Cluster Radius = 8.7\\
%     \hline   
%     \textit{typing on a keyboard} & \textit{bowling ball} \\
%  	\textit{Typing on keyboard} & \textit{electric tube}\\
%   	\textit{typing} & \textit{PVC pipe building pressure and release} \\
%     \textit{typing} & \textit{Error message on computer}\\
%     \textit{Typing on a keyboard} & \textit{High Speed Frisbee}\\
%     \textit{someone typing} & \textit{beer mug sliding on bar}\\
%     \textit{Someone typing on keyboard} & \textit{driving a car}\\
%     \textit{keyboard} & \textit{toy car hitting wall}\\
%     \textit{typing} & \textit{filling up a tub}\\
%     \textit{...} & \textit{...} \\
%   \hline  
%   \multicolumn{2}{l}{} \\ [1.2em] 
%   \end{tabular}
%   }
%   }
%   \label{example_labels}
%   \caption{Example responses from our free-text caption.  We see one sound that was very unambiguous and familiar (typing), and one sound which was intentionally modified from an already ambiguous sound (chair scraping on the floor).  We see a large divergence in sound source attribution for the ambiguous sound, which is captured in the cluster radius metric.}
  
% \end{table}
\end{floatrow}
\caption{Average ConceptNet embedding where the radius represents our $H_{cu}$ metric; red bubbles and the `\_mod' suffix are used to indicate sounds that have been intentionally modified. }
\label{cluster}
\end{figure*}

\subsection{Sourcing the Sounds}

All sounds in the dataset are sourced from the Freesound archive (\url{https://freesound.org}).  We built tools to rapidly explore the archive and re-label sound samples, searching for likely candidates based on tags and descriptions, and finally filtering by star and user ratings.  Each candidate sound was split into 5 second increments (and shorter sounds were extended to 5 seconds) during audition.  

A major goal in our curation was to find audio samples that spanned the space from "common and easy to identify" to "common but difficult to identify" and finally to "uncommon and difficult to identify".  We explicitly sought an even distribution of sounds in each broad category (approximately 130 sounds) using rudimentary blind self-tests. In sourcing sounds for the first two categories, we attempted to select samples that form common scenes one might encounter, such as \textit{kitchen, restaurant, bar, home, office, factory, airport, street, cabin, jungle, river, beach, construction site, warzone, ship, farm,} and \textit{human vocalization}.  We avoided any samples with explicit speech. 

To source unfamiliar/ambiguous sounds, we include a handful of digitally synthesized samples in addition to artificially manipulated everyday sounds.  Our manipulation pipeline applies a series of random effects and transforms to our existing samples from the former categories, from which we curated a subset of sufficiently unrecognizable results.  Effects include reverberation, time reversal, echo, time stretch/shrink, pitch modulation, and amplitude modulation.  

\subsection{Annotated Features}
\label{sec:pagestyle}

We began by designing an Amazon Mechanical Turk (AMT) experiment in which participants were presented with a sound chosen at random. Upon listening as many times as they desired, they then provided a free-text description alongside likert ratings of its familiarity, imageability, arousal, and valence (as depicted by the commonly used self-assessment manikins \cite{bradley2007international}).  The interface additionally captured metadata such as the time taken by each participant to complete their responses, the number of times a given sound was played, and the number of words used in the free-text response.  Roughly 12000 data points were collected through the experiment, resulting in approximately 30 evaluations per sound after discarding outliers (individual workers whose overall rankings deviate strongly from the global mean/standard deviation).  A reference screenshot of the interface and its included questions can be found at \url{github.com/mitmedialab/HCU400}.

\subsection{Audio Features}

Low level features were extracted using the Google VGGish audio classification network, which provides a 128-dimensional embedded representation of audio segments from a network trained to classify 600 types of sound events from YouTube \cite{VGGish}.  This is a standard feature extraction tool, and used in prominent datasets.  A comprehensive set of standard features extracted using the OpenSMILE toolkit \cite{opensmile} is also included.   

\subsection{Semantic Features}
\label{sec:typestyle}

A novel contribution of this work is the automation and extension of $H_{cu}$ using word embeddings and knowledge graphs.  Traditionally, these are used to geometrically capture semantic word relationships; here, we leverage the "clustering radius" of the set of label embeddings as a metric for each sound's $H_{cu}$.   

We employed three major approaches to embed each label: (1) averaging all constituent words that are nouns, verbs, adjectives, and adverbs-- a common/successful average encoding technique \cite{huang2017learning}-- (2) choosing only the first or last noun and verb, and (3) choosing a single 'head word' for each embedding based on a greedy search across a heavily stemmed version of all of the labels (using the aggressive Lancester Stemmer \cite{chris1990another}).  In cases where words are out-of-corpus, we auto-correct their spelling, and/or replace them with a synonym from WordNet where available \cite{wordnet}.  Labels that fail to cluster are represented by the word with the smallest distance to an existing cluster for that sound (using WordNet path-length).  This greedy search technique is used to automatically generate the group of labels used in the $H_{cu}$ calculation.  Both Word2Vec \cite{word2vec} and Conceptnet Numberbatch \cite{conceptnetnumberbatch} were tested to embed individual words.  %Word2Vec embeds words that appear in similar natural language contexts close together, while Conceptnet relies on structured semantic relations ('is-a', 'found-in', 'located-near', 'has', etc.) to generate embeddings. 

After embedding each label, we derived a 'cluster radius' score for the set of labels, using the mean and standard deviation of the distance of each label from the centroid as a baseline method. We also explore (k=3) nearest neighbor intra-cluster distances to reduce the impact of outliers and increase tolerance of oblong shapes.  Finally, we calculate the sum of weighted distance from each label subgroup to the largest 'head word' cluster-- a technique which emphasizes sounds with a single dominant label.  %Since this approach doesn't require a centroid, we include WordNet \cite{wordnet}-- a lexical graph that semantically links synonym subgraphs -- to find the distances between words in addition to Word2Vec and ConceptNet.  

We also include a location-based embedding to capture information pertaining to the likelihood of concept co-location in a physical environment.  In order to generate a co-location embedding, we implement a shallow-depth crawler that operates on ConceptNet's location relationships ('Located-Near', 'Located-At', etc) to create a weighted intersection matrix of the set of unique nouns across all our labels as a pseudo-embedding. Again, we derive the centroid location and mean deviation from the centroid of the labels (represented by the first unique noun) for a given sound sample.

Given the number of techniques, we compare and include only the most representative pipelines in our dataset.  All clustering approaches give a similar overall monotonic trend, but with variations in their derivative and noise.  Analysis of cluster labels in conjunction with scores suggests that a distance-from-primary-cluster definition is most fitting. Most embedding types are similar, but we prefer ConceptNet embeddings over others because it is explicitly designed to capture meaningful semantic relationships.

Our clustering results from a Processed ConceptNet embedding are plotted in Figure \ref{cluster}.  Intentionally modified sounds are plotted in red, and we see most sounds with divergent labeling fall into this category.  Sounds that have not been modified are in other colors-- here we see examples of completely unambiguous sounds, like human vocalizations, animal sounds, sirens, and instruments.

\section{Baseline Analysis and Discussion}
\label{sec:majhead}

\begin{figure}[h]
	\includegraphics[width=4.15cm]{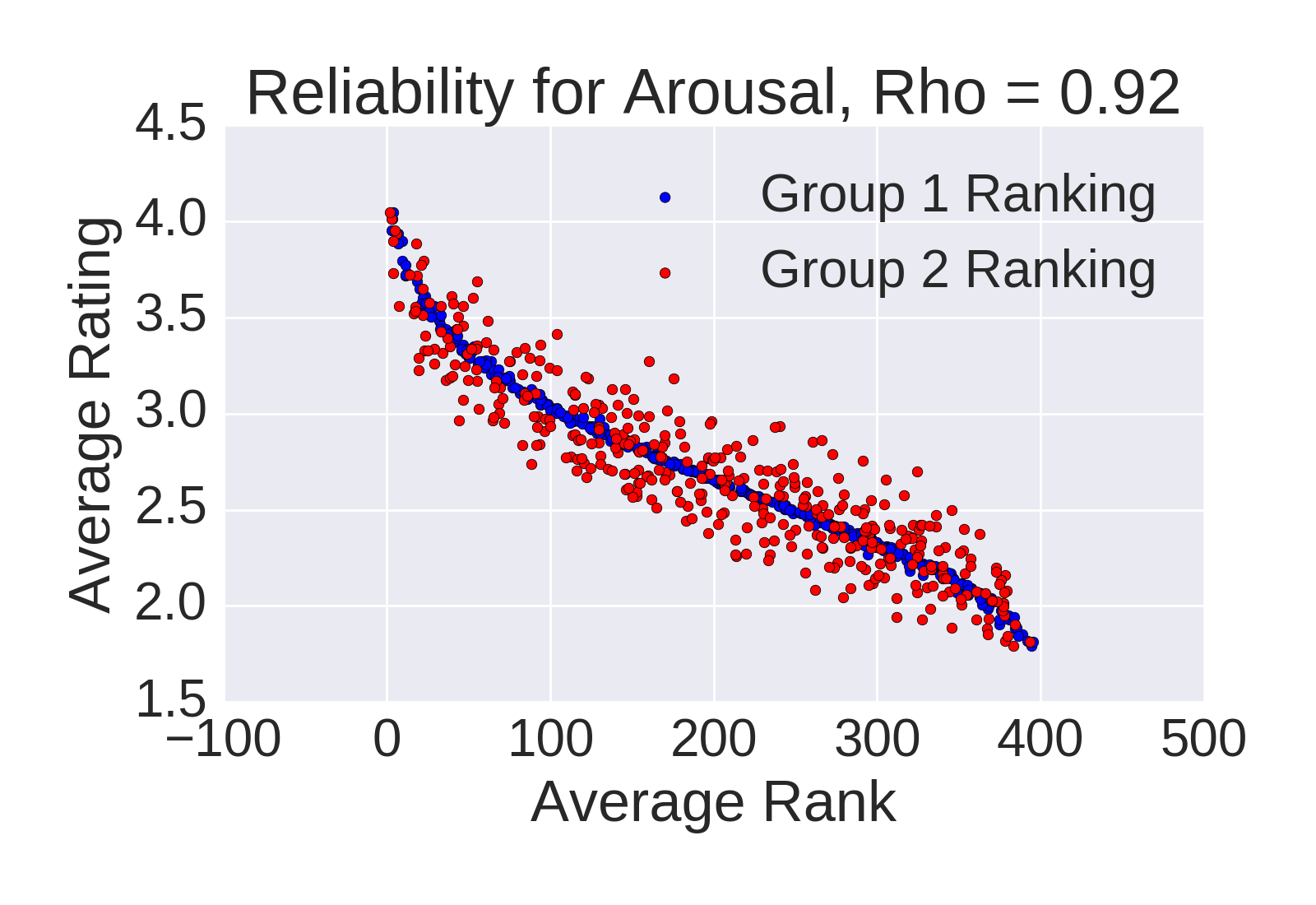}
    \includegraphics[width=4.15cm]{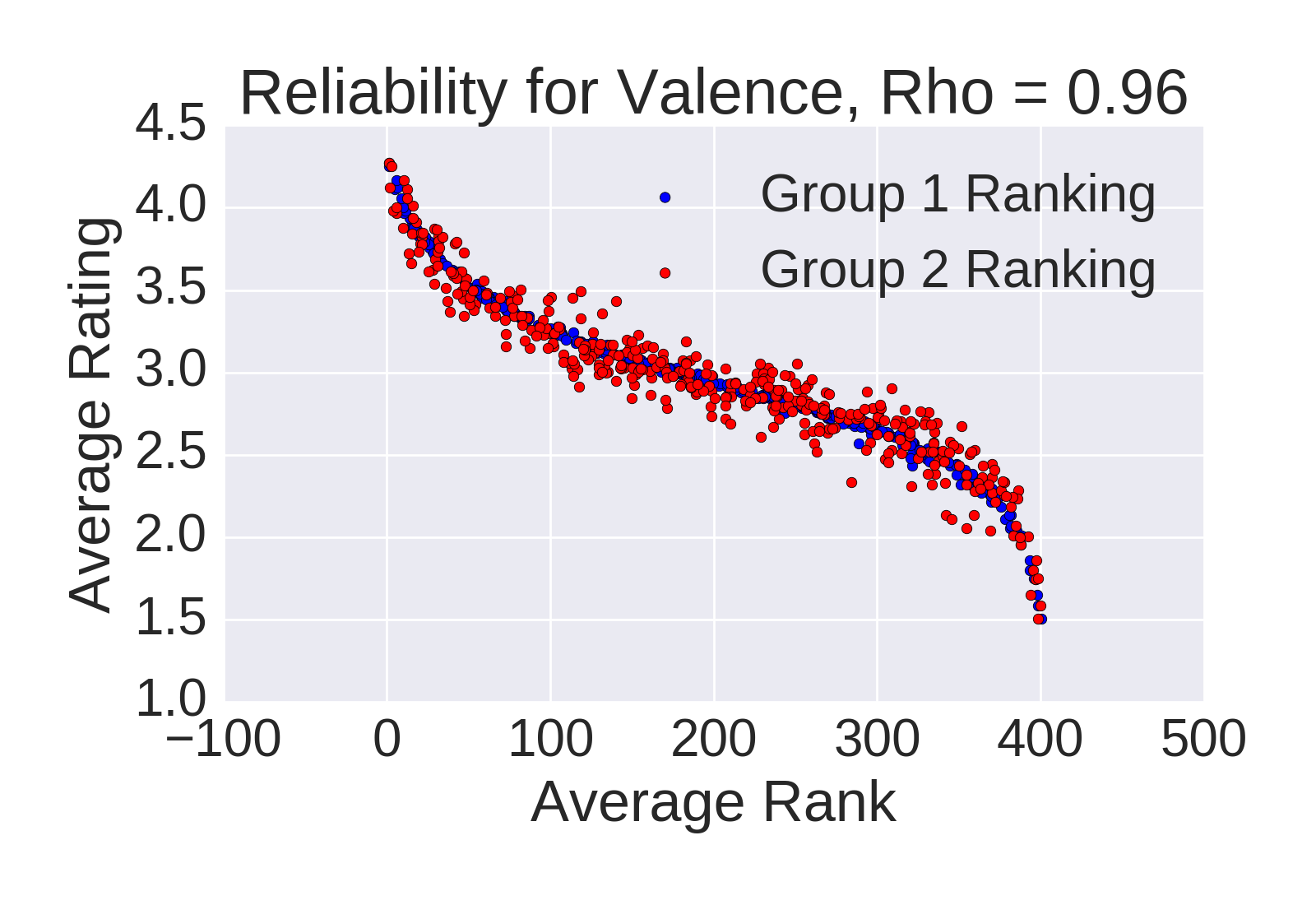}
    \includegraphics[width=4.15cm]{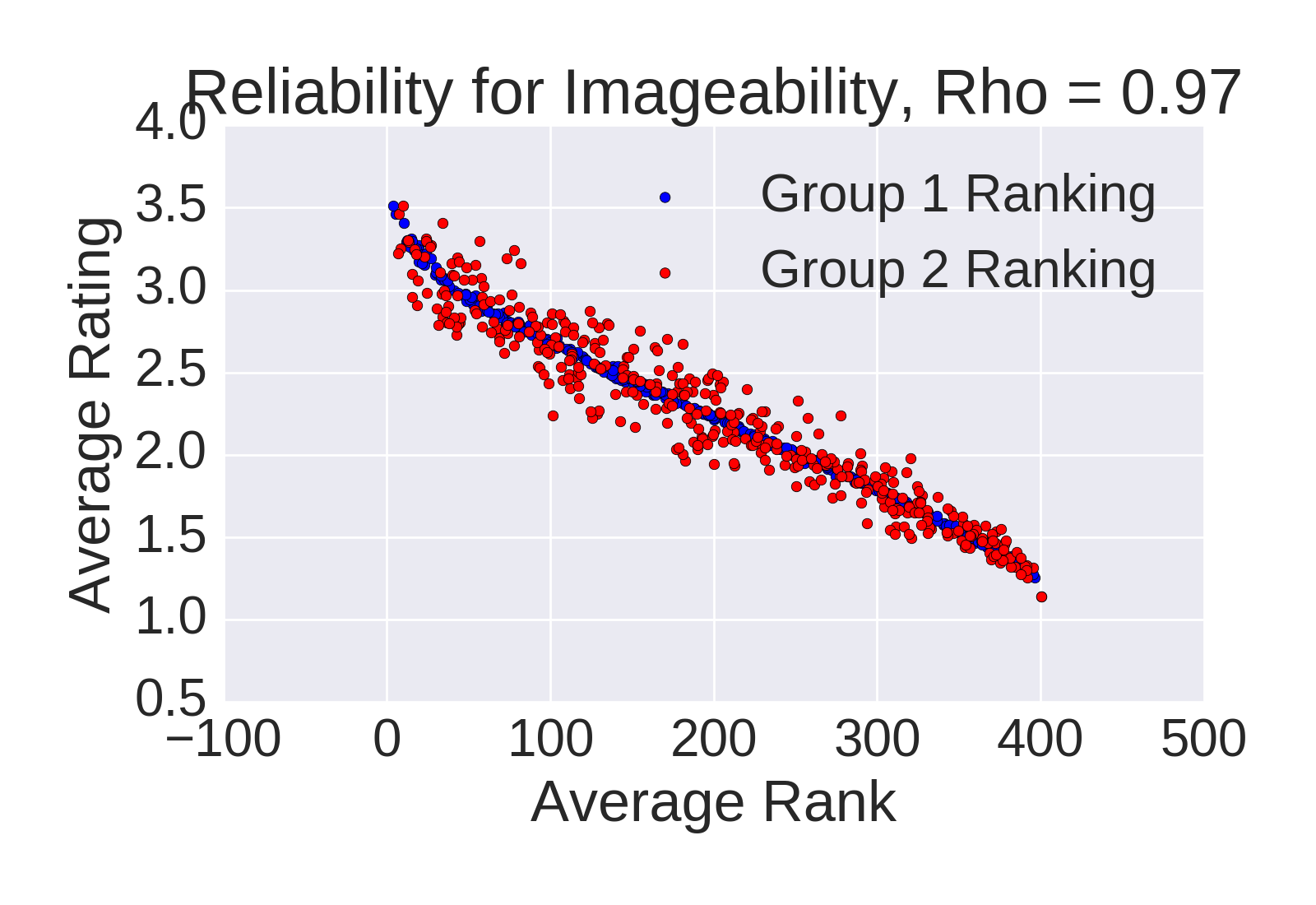}
    \includegraphics[width=4.15cm]{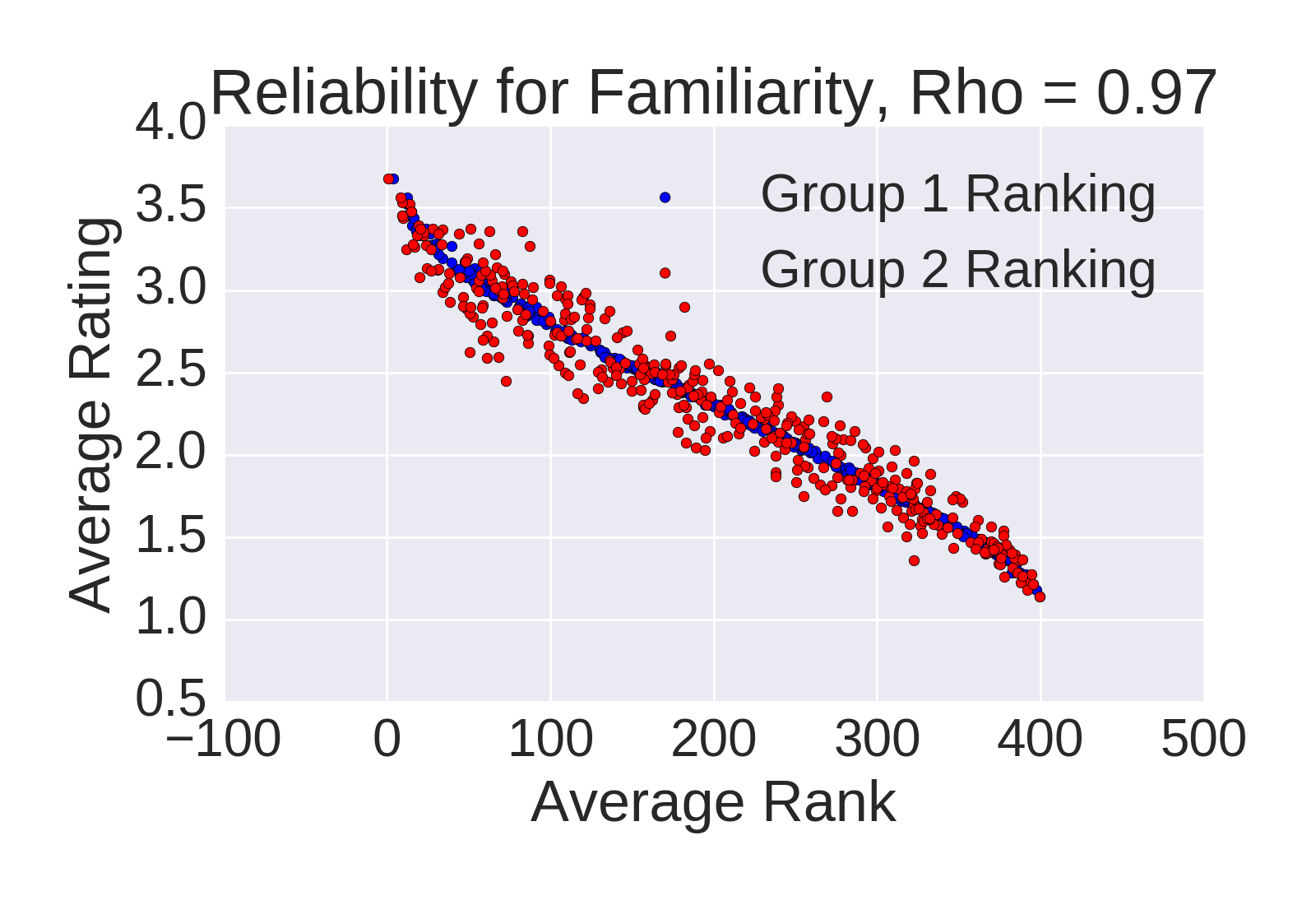}
    \caption{Split ranking correlation plots and Spearman rank coefficient values for the four likert annotated features.}
    \label{split}
\end{figure}

%\subsection{Feature Relationships and Discussion}

% \begin{figure}[!b]
	
% 	\includegraphics[angle=-90,origin=c,width=1.6cm]{final_shift_plots/v2_image_fam_mean.png}
%     \includegraphics[angle=-90,origin=c,width=1.6cm]{final_shift_plots/v2_image_fam_std.png}
%     \includegraphics[angle=-90,origin=c,width=1.6cm]{final_shift_plots/v2_val_ar_mean.png}
%     \includegraphics[angle=-90,origin=c,width=1.6cm]{final_shift_plots/v2_val_ar_std.png}
%     \centering
%     \includegraphics[angle=-90,origin=c,width=1.6cm]{final_shift_plots/v2_time_v_cluster.png}
%     \caption{Feature distributions grouped by extremes in the "Processed CNET" cluster metric; red points represent data at the 15th percentile or less; blue dots are 85th percentile and greater.}
%     \label{shift_plots}
% \end{figure}

\begin{figure*}[h!]

	\includegraphics[width=5.5cm]{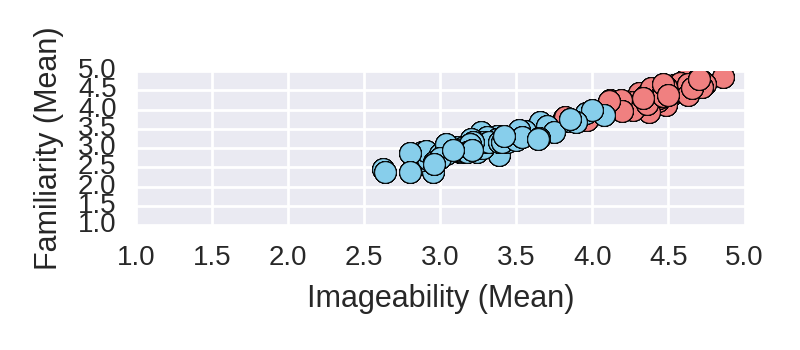}
    \includegraphics[width=5.5cm]{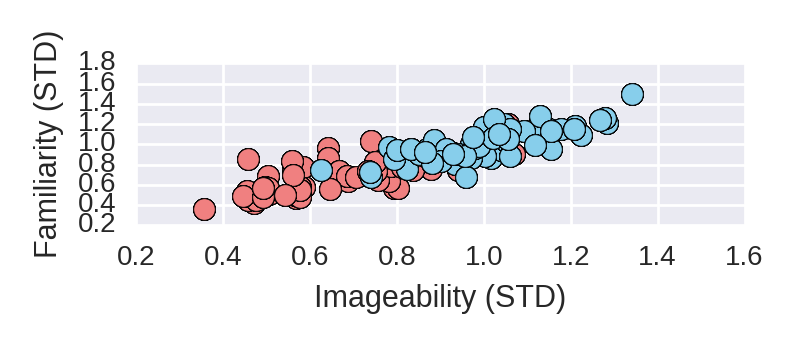}
    \includegraphics[width=5.5cm]{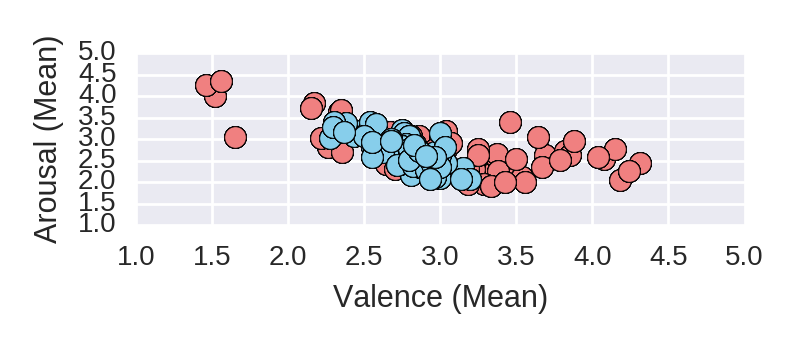}
    \includegraphics[width=5.5cm]{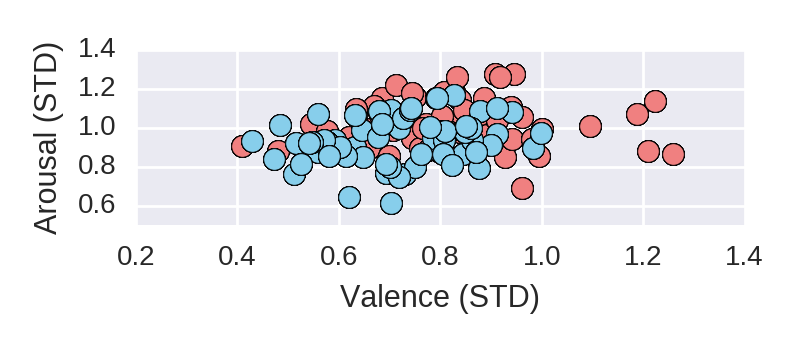}
    % \centering
    \includegraphics[width=5.5cm]{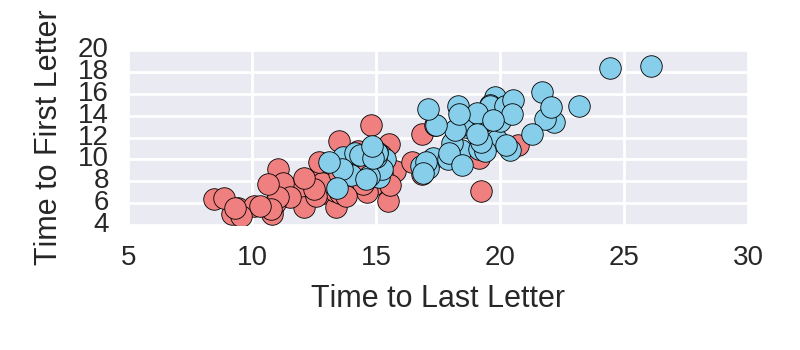}
    \caption{Feature distributions grouped by extremes in the "Processed CNET" cluster metric; red points represent data at $\leq$ 15th percentile (the most labeling agreement and least ambiguous); blue dots are $\geq$ 85th percentile (high $H_{cu}$).}
    \label{shift_plots}
\end{figure*}

\begin{figure}[h!]
	\includegraphics[width=8.5cm]{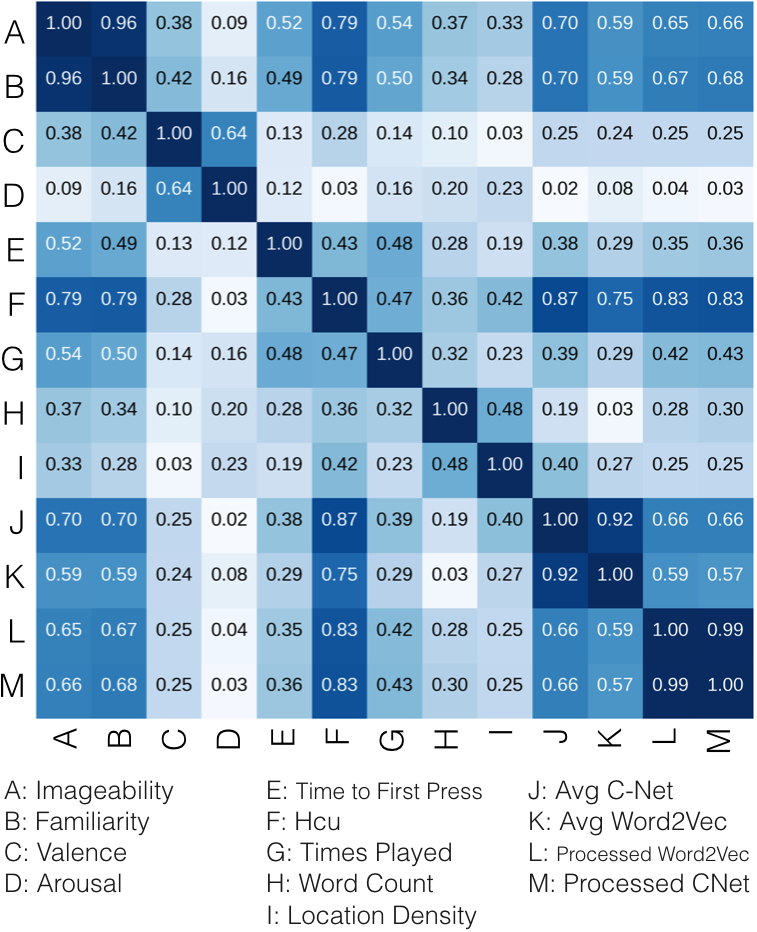}
    \caption{Correlation Matrix displaying the absolute value of the Pearson correlation coefficient between the mean values of annotated features, metadata, and four representative word embedding based clustering techniques.}
    \label{matrix}
\end{figure}

First, we find that the likert annotations are reliable amongst online workers, using a split ranking evaluation adapted from \cite{faces}. Each of the groups consisted of 50 \% of the workers, and the mean ranking was computed after averaging N=5 splits.  The resulting spearman rank coefficient value for each of the crowd-sourced features is given in Figure \ref{split}.  This provides the basis for several intuitive trends in our data, as shown by Figure \ref{matrix} -- we find a near linear correlation between imageability and familiarity, and a significant correlation between arousal and valence. We also find a strong correlation between imageability, familiarity, time-based individual measures of uncertainty (such as such "time to first letter" or "num of times played"), and the label-based, aggregate measures of uncertainty (the cluster radii and $H_{cu}$). 

We next see strong evidence of the value of word embeddings as a measure of causal uncertainty -- the automated technique aligns well with the split of modified/ non-modified sounds (see Fig. \ref{cluster}) and a qualitative review of the data labels.  Our measure also goes one step beyond $H_{cu}$, as the cluster centroid assigns representative content to the group of labels.  Initial clustering of sounds by their embedded centroids reveals a relationship between clusters and emotion rankings when the source is unambiguous, which could be generalized to predict non-annotated sounds (i.e., sirens, horns, and traffic all cluster together and have very close positive arousal and negative valence rankings; similar kinds of trends hold for clusters of musical instruments and nature sounds).

Furthermore, we use this data to explore the causal relationship between average source uncertainty and individual assessment behavior.  In Figure \ref{shift_plots}, we plot the distributions of pairs of features as a function of data points within the 15th (red) and greater than 85th (blue) percentile of a single cluster metric ("Processed CNET").  It confirms a strong relationship between the extremes of the metric and individual deliberation (bottom right), as reported by \cite{ballas1993common}.  We further find that more ambiguous sounds have less extreme emotion ratings (top right); the data suggest this is not because of disagreement in causal attribution, but because individuals are less impacted when the source is less clear (bottom left).  This trend is not true of imageability and familiarity, however; as sounds become more ambiguous, individuals are more likely to diverge in their responses (top center).  Regardless, we find a strong downward trend in average familiarity/imageability scores as the source becomes more uncertain (top left). 

\section{Conclusion}

It is known that aural phenomenology rests on a complex interaction between a presumed sound source, the certainty of that source, the sound's acoustic features, its ecological frequency, and its familiarity.  We have introduced the HCU400-- a dataset of everyday and intentionally obscured sounds that reliably captures a full set of affective features, self-reported cognitive features, timing, and free-text labels.  We present a new technique to quantify $H_{cu}$ using the distances between word embeddings of free text labels.  Our analysis demonstrates (1) the efficacy of a quantified approach to $H_{cu}$ using word embeddings; (2) the quality of our crowd-sourced likert ratings; and (3) the complex relationships between global uncertainty and individual rating behavior, which offers novel insight into our understanding of auditory perception.

\bibliographystyle{IEEEbib}
\bibliography{strings,refs}

\end{document}